\documentclass[sigconf]{acmart}

\setcopyright{none}
\renewcommand\footnotetextcopyrightpermission[1]{}
\acmConference{}{}{}\acmYear{}\copyrightyear{}\acmBooktitle{}\acmISBN{}\acmDOI{}

\usepackage[normalem]{ulem}
\usepackage{graphicx}
\usepackage{microtype} 
\usepackage{xurl}
\usepackage{longtable}

\AtBeginDocument{%
  \providecommand\BibTeX{{%
    \normalfont B\kern-0.5em{\scshape i\kern-0.25em b}\kern-0.8em\TeX}}}

\begin{document}

\title{A Virtual Laboratory for Managing Computational Experiments}

\author{Eleni Adamidi}
\email{eleni.adamidi@athenarc.gr}
\orcid{1234-5678-9012}
\authornotemark[1]
\affiliation{%
  \institution{Information Management Systems Institute}        
  \institution{Athena Research Center}
  \city{Athens}
  \country{Greece}
}

\author{Panayiotis Deligiannis}
\email{deligianp@athenarc.gr}
\affiliation{%
  \institution{Information Management Systems Institute}        
  \institution{Athena Research Center}
  \city{Athens}
  \country{Greece}
}

\author{Nikos Foutris}
\email{nikos.foutris@athenarc.gr}
\affiliation{%
  \institution{Information Management Systems Institute}        
  \institution{Athena Research Center}
  \city{Athens}
  \country{Greece}
}

\author{Thanasis Vergoulis}
\email{vergoulis@athenarc.gr}
\affiliation{%
  \institution{Information Management Systems Institute}        
  \institution{Athena Research Center}
  \city{Athens}
  \country{Greece}
}

\renewcommand{\shortauthors}{Adamidi, et al.}

\begin{abstract}

Computational experiments have become essential for scientific discovery, allowing researchers to test hypotheses, analyze complex datasets, and validate findings. However, as computational experiments grow in scale and complexity, ensuring reproducibility and managing detailed metadata becomes increasingly challenging, especially when orchestrating complex sequence of computational tasks. To address these challenges we have developed a virtual laboratory called SCHEMA lab, focusing on capturing rich metadata such as experiment configurations and performance metrics, to support computational reproducibility. SCHEMA lab enables researchers to create experiments by grouping together multiple executions and manage them throughout their life cycle. In this demonstration paper, we present the SCHEMA lab architecture, core functionalities, and implementation, emphasizing its potential to significantly enhance reproducibility and efficiency in computational research.

\end{abstract}


\maketitle
\section{Introduction}
Computational experiments are essential for advancing scientific discovery, enabling researchers to test hypotheses, analyze data, and simulate complex systems. The growing complexity of these experiments, however, presents challenges related to reproducibility and metadata management. Precise documentation and structured execution environments have become essential to ensuring experiments can be consistently reproduced and verified across diverse computing infrastructures. As highlighted by Sandve et al. \cite{Sandve2013} and Peng \cite{Peng2011}, computational reproducibility demands meticulous management of workflows, configurations, and performance data.

Managing the increasing scale and complexity of scientific workflows requires advanced tools capable of automating and tracking numerous computational tasks simultaneously. Researchers frequently face difficulties in reliably reproducing experiments due to inconsistencies in execution environments, incomplete metadata documentation, and lack of structured management practices. Such challenges significantly hinder scientific transparency, verification, and collaboration.

Containerization technologies have fundamentally transformed computational research by offering consistent and portable execution environments. Containerization, particularly through technologies such as Docker \cite{Boettiger2015} has enabled researchers to encapsulate applications along with their dependencies, thereby simplifying deployment across diverse systems and mitigating issues related to software configuration. Moreover, existing systems like Galaxy \cite{Afgan2016} and Nextflow \cite{DiTommaso2017} have made significant strides in automating scientific workflows.  While these advances facilitate reliable execution of computational tasks and workflows, environments that enable researchers to create and holistically manage complex computational experiments-comprising of multiple containerized tasks, rich metadata capture, and reproducibility considerations-are needed.

SCHEMA lab\footnote{
SCHEMA lab: \url{https://github.com/athenarc/schema-lab},
} addresses this gap by providing an open source virtual laboratory that empowers researchers to create, manage, and monitor containerized computational experiments with ease. Building upon previous work \cite{Vergoulis2021}, SCHEMA lab allows users to submit and track the execution of both individual tasks and complex workflows while capturing essential metadata for performance analysis and reproducibility.

In this demonstration paper, we present the system architecture, core functionalities, and implementation of SCHEMA lab, illustrating its potential for enhancing computational reproducibility and experiment management.

\section{Background and Related work}
Over the past decade, significant advances have been made in container orchestration, workflow management, and reproducibility practices within computational research. Many platforms—such as Galaxy \cite{Afgan2016}, Snakemake \cite{Koster2012}, and Nextflow \cite{DiTommaso2017}—have been developed to coordinate multi-step scientific workflows with varying degrees of automation and user support. These systems excel in orchestrating interdependent tasks, yet researchers may benefit from additional flexibility in managing diverse computational workloads, particularly when systematically capturing detailed metadata to enhance reproducibility and experiment documentation.

SCHEMA lab addresses this need by providing a flexible, lightweight environment for managing containerized computational tasks. It supports the rapid execution of standalone tasks—ideal for quick tests and prototyping—while also enabling users to dynamically group tasks into larger experiments. When tasks are run as part of a grouped experiment, SCHEMA lab aggregates detailed information on experiment configuration, performance metrics, and resource consumption, thereby facilitating easier documentation and reproducibility. In contrast, executing standalone tasks focuses on rapid iteration and debugging.

Reproducibility remains a central theme in computational research. Prior studies (e.g., Sandve et al. \cite{Sandve2013} and Peng \cite{Peng2011}) have underscored the importance of meticulously documenting computational processes to enable the exact replication of results. In this context, recent advances in data packaging standards—such as the RO-Crate framework \cite{Sefton2020}—offer promising strategies for encapsulating experimental data and metadata in a standardized, shareable format.

By aligning with established paradigms in containerization, workflow management, and reproducibility, SCHEMA lab is positioned as a tool that not only executes tasks efficiently but also adapts to the evolving needs of researchers—whether they require rapid prototyping or comprehensive experiment management.

\section{System Overview}
The architecture of SCHEMA lab is designed to provide a seamless and flexible environment for managing containerized computational experiments. The system is divided into two primary components: the SCHEMA lab front-end and the SCHEMA api back-end. Together, these components enable researchers to submit, execute, monitor, and manage containerized tasks and computational experiments efficiently. The following subsections describe the overall architecture, the front-end functionalities, the back-end services, and the integration between these components.

\subsection{High-Level Architecture}
Our platform supports the submission and monitoring of containerized task execution requests. Its purpose is to act as a gateway between users and the task execution environment, performing necessary authentication and authorization checks, recording submitted task requests, and scheduling valid, authorized tasks for execution via exposed RESTful endpoints.

Under the hood SCHEMA api works over a Kubernetes cluster that runs tasks shipped in Docker containers. Rather than communicating directly with Kubernetes, SCHEMA api schedules task executions through TESK—an implementation of the Task Execution Service (TES) API \cite{tesapi} developed under the Global Alliance for Genomics and Health (GA4GH) \cite{ga4gh}. Consequently, SCHEMA api requires an operational deployment of TESK \cite{tesk} that is accessible via HTTP.

On the front-end, SCHEMA lab communicates with SCHEMA api using RESTful calls, ensuring that user actions (such as task or workflow submission and experiment creation) are reliably forwarded to the back-end. The resulting execution status and metadata are then pushed back to SCHEMA lab, enabling real-time monitoring and control of computations. 

\begin{figure}[ht]
  \centering
  \includegraphics[width=0.6\linewidth]{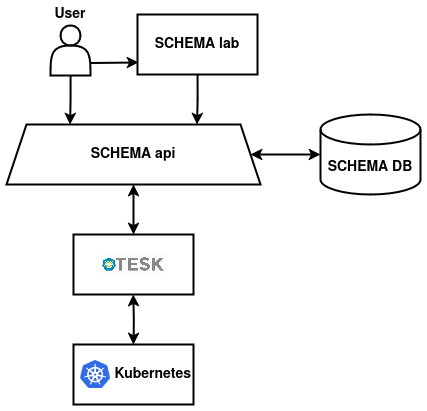}
  \caption{High-Level Architecture of SCHEMA api. This diagram illustrates how SCHEMA api acts as a gateway between users and the containerized execution environment by performing authentication, scheduling containerized tasks via TESK, and providing monitoring endpoints.}
  \Description{A diagram showing the high-level architecture of SCHEMA api, including its interaction with TESK for task scheduling on a Kubernetes cluster.}
  \label{fig:high_level_architecture}
\end{figure}

A high-level diagram of this architecture and its dependencies is depicted in Figure~\ref{fig:high_level_architecture}. In essence, SCHEMA api preserves the logical abstraction of tasks, contexts, and experiments, while TESK handles the low-level container scheduling on Kubernetes.

\subsection{Back-End: SCHEMA api}

SCHEMA api is the engine that powers the admission and scheduling of containerized tasks. At its core, it is designed to orchestrate the execution of containerized software, available in any reachable registry, as well as passing to the execution input files and storing produced output files. Within the SCHEMA api ecosystem, several resources are used to achieve this functionality:

\begin{itemize}
    \label{schema-api/resources-list}
    \item \textbf{Tasks:} single-job containerized executions
    \item \textbf{Workflows:} multi-job containerized executions with known intra-job dependencies
    \item \textbf{Contexts:} sets of executions which can be ran by certain users with certain quotas
    \item \textbf{Quotas:} optional numerical limits that control the submitted executions
    \item \textbf{Experiments:} sets of selected executions that are grouped together and represent a computational effort
\end{itemize}

\begin{figure}[ht]
  \centering
  \includegraphics[width=\linewidth]{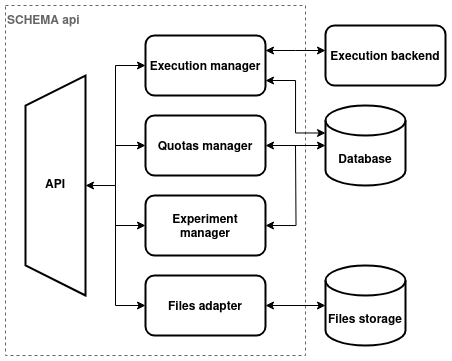}
  \caption{Internal architecture of SCHEMA-api}
  \Description{A diagram illustrating the internal architecture of SCHEMA api.}
  \label{fig:internal-architecture}
\end{figure}

SCHEMA api is organized into several components: the \emph{API}, the \emph{execution manager}, the \emph{quotas manager}, the \emph{experiments manager} and the \emph{files adapter}. Furthermore, it uses these components in order to manage external systems and data stores like the \emph{execution backend}, a \emph{relational database} and a \emph{files storage} holding the files and directories used for input and output in containerized computations. This architecture is depicted in Figure~\ref{fig:internal-architecture}.

\subsubsection{Execution manager}\label{schema-api/execution-manager}

The execution manager materializes the core functionality of managing submitted workflows in SCHEMA api. When a new execution request is submitted, it performs preliminary steps like generating necessary execution metadata or resolving workflow job execution order. Moreover, it evaluates the effective quotas based on the submitting user and the execution context. If the execution can run, based on the current context and user resource utilization, it stores execution information in the database and finally schedules the execution on a supported execution backend.

\begin{figure}[ht]
  \centering
  \includegraphics[width=0.9\linewidth]{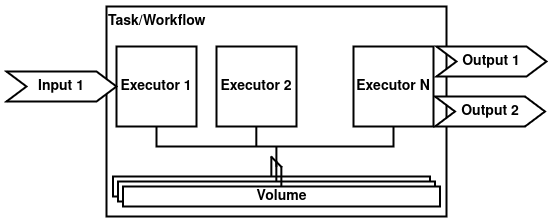}
  \caption{Overview of an indicative workflow with one input, two outputs and three volumes}
  \Description{A diagram depicting the representation of an execution in SCHEMA api, illustrating an execution as a parent entity with multiple executors, each having its own configuration, and associated input/output mount points and volumes.}
  \label{fig:task_module}
\end{figure}

While SCHEMA api is designed to be easily extended for multiple execution backends, in its production deployment it talks directly to TESK, which implements the TES API. In fact, SCHEMA api is inspired by TES API in the way it represents task and workflow data for the data schema and the communication with TESK. In essence, each task or workflow consists is described with the following core entities, which are also illustrated in Figure~\ref{fig:task_module}: 
\begin{itemize} 
    \item \textbf{Task/Workflow:} The primary execution entity, which aggregates the configuration for that corresponds to a task or workflow. 
    \item \textbf{Executors:} Each execution can have multiple executors, each corresponding to a single containerized job. This allows complex computations across multiple containerized images. 
    \item \textbf{Environment Variables:} Executors can be configured with environment variables that are applied in their respective containers. 
    \item \textbf{Input/Output Mount Points:} Mappings that move files into the container for the first executor and pull files from the container of the last executor. 
    \item \textbf{Volumes:} Shared directories accessible by all executors, facilitating data sharing between different stages of a multi-job (workflow) execution. 
\end{itemize}

\subsubsection{Quotas manager} In order to control the execution footprint on resources and fair sharing, SCHEMA api allows for the definition of quotas that enforce limits on user and context resources. These quotas are stored in the database and may be evaluated at different places in an execution's life-cycle. The quotas manager is responsible for managing these limits and computing the effective quotas for any new execution.

\subsubsection{Experiment manager}

The experiment manager allows users to create experiments that represent a computation effort. Additional users can be added in the experiment that are already co-participants with the experiment creator in an execution context. Tasks and workflow executions ran within the same execution context, by experiment participants, can then be associated with the created experiment.

\subsubsection{Files adapter} The files adapter communicates with an external S3 instance and manages the input and output files for each user, which are stored in separate user buckets. It can carry out basic file management operations like renaming/moving a file, deleting a file, listing directory contents, and retrieving file metadata. Moreover, it issues download and upload URLs that can be given back to users for access to their bucket storage.

\subsubsection{API}

The API component is responsible for exposing REST API endpoints for the management of the rest of the components. Incoming requests reaching the API are initially passed through an authorization mechanism that restricts access to users possessing an active API key. Authorized requests are then routed to the corresponding component, based on the requested API path. These handlers carry out data validation on the input data, trigger the necessary actions on the underlying components and finally respond to the received requests.

API endpoints can be organized into groups that correspond to actions over the resources introduced in Section~\ref{schema-api/resources-list}. The tasks endpoint group handles the submission and management of single-job containerized tasks. Workflows can be encoded with a native format, as described in Appendix~\hyperref[appendix:native-workflow-specification]{A}, and can be managed by similar API endpoints. Moreover, in the context of reproducibility, SCHEMA api exposes endpoints that allow users to manage experiments. Finally, there are auxiliary operations for file management. The endpoints for the above groups are further described in Appendix ~\hyperref[appendix:endpoints]{B}.
Additional technical details—including API specification, deployment instructions, and configuration guidelines—are provided in the SCHEMA api documentation.\footnote{
API specification: \url{https://schema.athenarc.gr/docs/schema-api/spec},\\
Deployment instructions: \url{https://schema.athenarc.gr/docs/schema-api/deployment},\\
Configuration guidelines: \url{https://schema.athenarc.gr/docs/schema-api/deployment/config}}

\subsection{Integration and Data Flow}
The integration between SCHEMA lab and SCHEMA api is designed to ensure reliable, real-time communication and efficient handling of computational tasks. The data flow encompasses several key stages that enable seamless interaction and comprehensive experiment management.

The process begins with \textbf{task submission}. A user initiates a new task via the SCHEMA lab interface, which triggers a RESTful request to SCHEMA api. Depending on the nature of the task, this may involve a \texttt{POST /api/tasks} request for standalone tasks or a \texttt{POST /api/workflows} request for grouped tasks that involve multiple dependent executions.

Upon receiving the submission, SCHEMA api handles \textbf{task scheduling and execution}. The API validates the request and stores it in a structured data store to maintain a record of the configuration and metadata. Once recorded, the task is scheduled for execution through TESK, which interfaces directly with a Kubernetes cluster. TESK manages the actual deployment of the containerized jobs, while SCHEMA api captures the initial status information and communicates it back to the front end, allowing users to monitor the progress of their tasks in real time.

Throughout execution, SCHEMA lab supports \textbf{status monitoring and data aggregation}. The front end periodically queries SCHEMA api (e.g., via \texttt{GET /api/tasks/\{uuid\}}) to fetch the latest status updates.

The system also incorporates \textbf{error handling and feedback mechanisms}. If execution errors occur or if a user triggers task cancellation through the \texttt{POST /api/tasks/\{uuid\}/cancel} endpoint, SCHEMA api provides logs for these events to the front end. Users receive detailed error messages facilitating rapid troubleshooting of failed executions.

Looking ahead, SCHEMA lab is designed with extensibility in mind. A planned enhancement is the integration of \textbf{RO-Crate export functionality}, which will allow experiments to be packaged with standardized data and metadata descriptions. This feature aims to further strengthen reproducibility and enable easier sharing and publication of computational experiments.

Overall, each stage of the workflow is supported by RESTful communication patterns and robust error-handling mechanisms. This design ensures that SCHEMA lab delivers a seamless user experience, even under complex computational loads and multi-tasking scenarios.

\subsection{Front-End: SCHEMA lab}

\begin{figure}[ht]
    \centering
    \includegraphics[width=\linewidth]{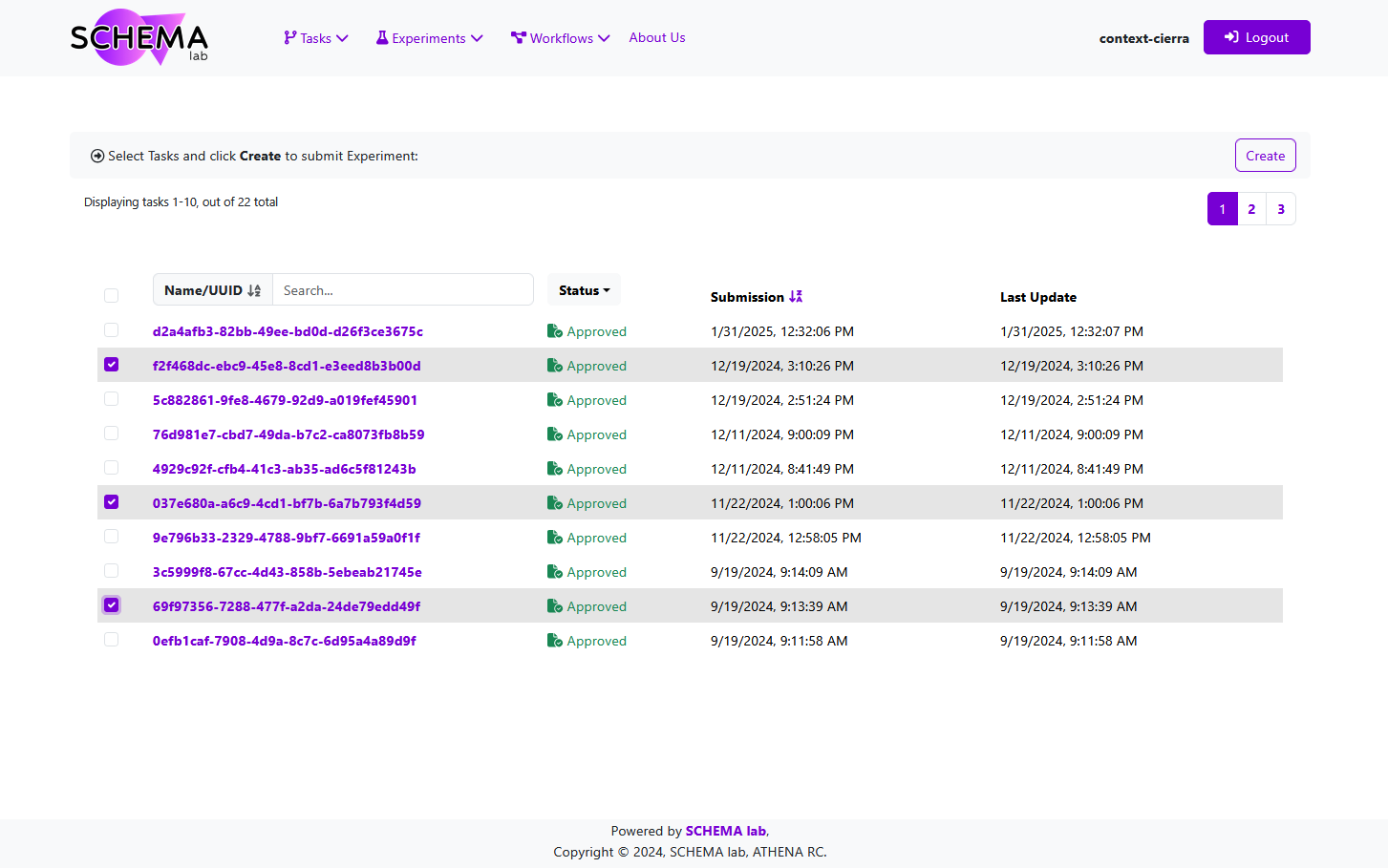}
    \caption{SCHEMA lab interface for creating an experiment by selecting executed tasks.}
    \Description{A screenshot of SCHEMA lab's UI, displaying a list of executed tasks with check-boxes, allowing users to select tasks and click 'Create' to form a new experiment.}
    \label{fig:experiments}
\end{figure}

The SCHEMA lab front-end provides an intuitive interface for users to interact with the system. Its main features include: \begin{itemize} \item \textbf{Task Submission and Monitoring:} Users can submit both standalone single containerized tasks and workflow tasks. Users can view all submitted tasks along with their unique identifiers, execution status, submission time, and last update timestamp. Figure~\ref{fig:task_dashboard} illustrates the task management interface in SCHEMA lab. The interface presents task execution details in a tabular format, with clear indicators for different statuses such as Approved, Running, Completed, Scheduled, and Error. The rightmost column includes interactive buttons that allow users to cancel or re-execute a task if necessary.

\begin{figure}[ht]
  \centering
  \includegraphics[width=\linewidth]{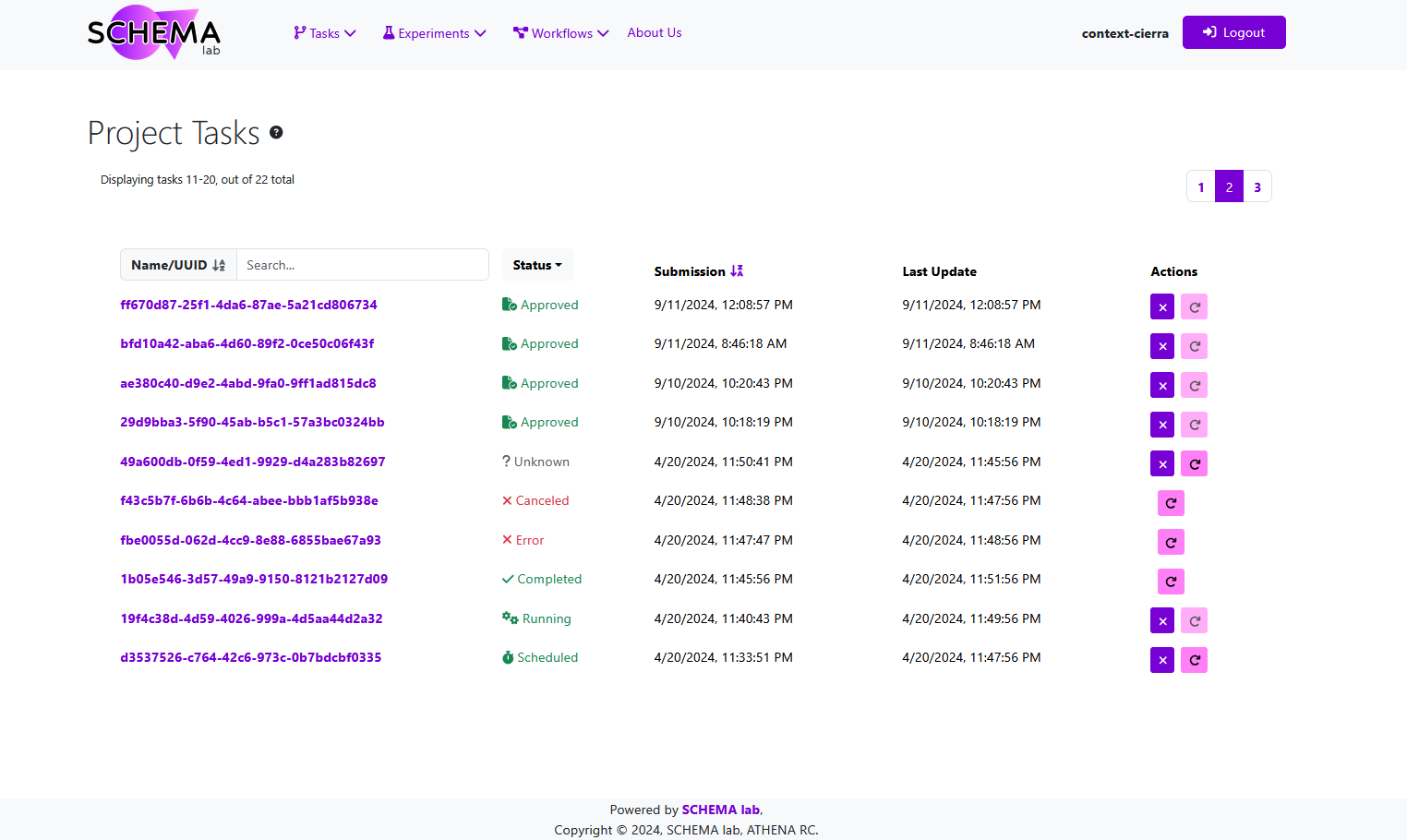}
  \caption{Task Management Interface in SCHEMA lab. Users can view and manage submitted tasks, track their statuses, and perform actions such as canceling or resubmitting executions.}
  \Description{A screenshot of the SCHEMA lab interface displaying a table of tasks with status labels and action buttons for user interaction.}
  \label{fig:task_dashboard}
\end{figure}

\item \textbf{Experiment Management:} The interface supports the grouping of executed single tasks or workflow tasks into experiments using interactive elements such as check-boxes. Users can create, update, and delete experiments, as well as review all the experiments they have created in a context. The UI provides a simple way to select completed tasks and create an experiment, as illustrated in Figure~\ref{fig:experiments}.
\end{itemize}

\section{Availability \& Resources}
SCHEMA lab and SCHEMA api are designed as open platforms to support containerized task and workflow execution and management and experiment creation. To facilitate adoption, development, and integration, we provide the following resources:

\begin{itemize}
    \item \textbf{GitHub Repositories:} The source code for both SCHEMA api and SCHEMA lab is openly available for contributions and issue tracking:
    \begin{itemize}
        \item SCHEMA api: \url{https://github.com/athenarc/schema-api/tree/main}
        \item SCHEMA lab: \url{https://github.com/athenarc/schema-lab}
    \end{itemize}
    
    \item \textbf{Swagger API Documentation:} Developers can explore the SCHEMA api endpoints and test API calls via the Swagger UI, available at \url{https://api.hypatia-comp.athenarc.gr/}.
    
    \item \textbf{Code and User Documentation:} The code and user documentation, including API specifications and deployment guides, is actively maintained and can be accessed at \url{https://schema.athenarc.gr}. This resource is continuously evolving to reflect the latest developments.
\end{itemize}

\section{Demonstration Scenarios}

During the demonstration session, we will showcase SCHEMA lab's functionalities using a deployment on the HYPATIA Cloud Infrastructure\footnote{https://hypatia.athenarc.gr/}, which leverages a large computational cluster. Specifically, we will illustrate SCHEMA lab's capabilities interactively by addressing scenarios defined in real-time by the audience. Additionally, we have prepared several illustrative scenarios designed to highlight key aspects and strengths of the platform:

\paragraph{Scenario 1: Submitting and Monitoring Individual Tasks}
In this scenario, we will demonstrate how users can quickly submit individual tasks and monitor their execution through SCHEMA lab. Users will see how tasks are submitted, their progress tracked in real-time, and outputs retrieved seamlessly via the platform's intuitive interface.

\paragraph{Scenario 2: Defining and Executing Workflows}
We will present SCHEMA lab's workflow orchestration capabilities, highlighting how users can define, execute, and monitor workflows consisting of multiple containerized tasks with interdependencies. This demonstration will illustrate the workflow creation process, execution monitoring, and management of dependencies within SCHEMA lab.

\paragraph{Scenario 3: Creating Computational Experiments with Multiple Tasks or Workflows}
This scenario will demonstrate the platform's functionality for creating comprehensive computational experiments by selecting and combining multiple tasks or workflows. Attendees will have the opportunity to explore the system hands-on and test any scenario of their choice.

\section{Conclusion}
SCHEMA lab, powered by SCHEMA api, provides a comprehensive and user-friendly platform for executing and managing containerized computational tasks. Designed to address the complexities of scientific experiments, it enables users to submit, monitor, and manage both individual containerized tasks and workflows efficiently. Through its integration with TESK and Kubernetes, the system supports scalable, distributed execution while maintaining a high level of control and reproducibility.

One of the key contributions of SCHEMA lab is its ability to bridge the gap between containerized task execution and structured experiment management. Unlike traditional workflow engines, SCHEMA lab not only supports individual executions but also allows users to group multiple executions into experiments, providing a higher-level abstraction for computational research. This design enhances the traceability and reproducibility of scientific computations, making it easier for researchers to document, share, and verify their results. The system also offers a web-based interface that simplifies interaction for users, providing an intuitive way to manage and inspect tasks without requiring deep technical expertise.

Moreover, the SCHEMA api backend offers a modular and extensible architecture through secure RESTful endpoints for task, workflow, and experiment management. This design fosters interoperability, facilitating future integration with other scientific computing platforms.

To further enhance the capabilities of SCHEMA lab and SCHEMA api, several future developments are planned. First, we intend to integrate support for additional workflow languages—such as Nextflow, enabling users to leverage advanced workflow paradigms and tap into the rich ecosystem of existing scientific workflow tools. Secondly, we plan to implement RO-Crate export functionality, which will allow experiments to be packaged in a standardized, lightweight format that encapsulates both data and metadata, thereby facilitating reproducibility and data sharing across platforms. For this purpose we plan to use a custom RO-Crate profile focused in performance metadata \cite{Adamidi2024}. Finally, we plan to establish integrations with external repositories, such as RO-Hub, and Workflow Hub. connecting with Workflow Hub to streamline the sharing and reuse of computational workflows

This demonstration has showcased how SCHEMA lab simplifies the orchestration of containerized tasks and workflows, making it an invaluable tool for researchers and organizations managing computational experiments. By prioritizing usability, scalability, and reproducibility, SCHEMA lab stands as a versatile and forward-thinking solution for modern scientific computing challenges.

\begin{acks}
TIER2 receives funding from the European Union's Horizon Europe research and innovation programme under grant agreement No 101094817. Views and opinions expressed are those of the author(s) only and do not necessarily reflect those of the European Union or the European Commission. Neither the EU nor the EC can be held responsible for them.
\end{acks}

\bibliographystyle{ACM-Reference-Format}
\bibliography{sample-base}

\appendix
\section*{APPENDIX A: NATIVE WORKFLOW SPECIFICATION}\label{appendix:native-workflow-specification}
SCHEMA api supports declarative workflows that are defined in a specific format. It introduces a native workflow specification that directly maps to internal data structures and can be easily serialized to JSON format. Although this specification is based on the task structure described in Section~\ref{schema-api/execution-manager}, it uses certain additional parameters that describe executor file dependencies and allow the resolution of the order of execution.

SCHEMA api enables the prospective integration of additional workflow languages by utilizing its native workflow specification. Extensions of SCHEMA api, aiming to support standard workflow languages, are planned to implement the necessary transpilation processes that generate the respective native workflow definitions. This allows internal SCHEMA api components, like the \emph{execution manager} to handle any workflow, regardless of its language. Therefore, SCHEMA api remains flexible by allowing the scheduling of the workflow execution to use either the original workflow language or the corresponding native definition.

Additional information regarding the support of workflows in SCHEMA api can be found at \url{https://schema.athenarc.gr/docs/schema-api/arch/workflows}.
\section*{APPENDIX B: SCHEMA API Endpoints}

\label{appendix:endpoints}

\subsection*{Task Endpoints}
\begin{itemize}
\item \texttt{POST /api/tasks}: Submit a new task execution request.
\item \texttt{GET /api/tasks}: Retrieve a list of submitted tasks.
\item \texttt{GET /api/tasks/\{uuid\}}: Get detailed information about a specific task.
\item \texttt{POST /api/tasks/\{uuid\}/cancel}: Cancel an ongoing task.
\item \texttt{GET /api/tasks/\{uuid\}/stdout}: Retrieve the standard output of a task.
\item \texttt{GET /api/tasks/\{uuid\}/stderr}: Retrieve the standard error of a task.
\item \texttt{GET /api/quotas}: Retrieve applied quotas for a user within a project.
\end{itemize}

\subsection*{Workflow Endpoints}
\begin{itemize}
\item \texttt{POST /api/workflows}: Submit a new workflow.
\item \texttt{GET /api/workflows}: List submitted workflows.
\item \texttt{GET /api/workflows/\{uuid\}}: Retrieve data of a specific workflow.
\item \texttt{POST /api/workflows/\{uuid\}/cancel}: Cancel a running workflow.
\item \texttt{GET /api/workflows/\{uuid\}/stdout}: Retrieve stdout of executed workflow jobs.
\item \texttt{GET /api/workflows/\{uuid\}/stderr}: Retrieve stderr of executed workflow jobs.
\end{itemize}

\subsection*{Experiment Endpoints}
\begin{itemize}
\raggedright
\item \texttt{GET /reproducibility/experiments}: List all experiments.
\item \texttt{POST /reproducibility/experiments}: Create a new experiment.
\item \texttt{GET /reproducibility/experiments/\{username\}/\{name\}}: Retrieve details of a specific experiment.
\item \texttt{PATCH /reproducibility/experiments/\{username\}/\{name\}}: Update an experiment.
\item \texttt{DELETE /reproducibility/experiments/\{username\}/\{name\}}: Delete an experiment.
\item \texttt{GET /reproducibility/experiments/\{username\}/\allowbreak\{name\}/\allowbreak tasks}: Retrieve tasks associated with an experiment.
\item \texttt{PUT /reproducibility/experiments/\{username\}/\allowbreak\{name\}/\allowbreak tasks}: Assign tasks to an experiment.
\end{itemize}

\subsection*{Storage Endpoints}
\begin{itemize}
\item \texttt{GET /storage/files}: List the files in the user's bucket.
\item \texttt{POST /storage/files}: Issue an upload link for a specific object.
\item \texttt{GET /storage/files/\{PATH\}}: Retrieve metadata or download link for a file.
\item \texttt{PATCH /storage/files/\{PATH\}}: Move or rename an object in the user's bucket.
\item \texttt{DELETE /storage/files/\{PATH\}}: Delete an object from the user's bucket.
\end{itemize}

\end{document}